\documentclass[aps,prl,twocolumn,groupedaddress]{revtex4}
\usepackage{graphicx}
\begin{document}

%\preprint{}

\title{Experimental observation of the `Tilting Mode' of an array of vortices
in a dilute Bose-Einstein Condensate}


\author{N. L. Smith, W. H. Heathcote, J. M. Krueger and C. J. Foot}
%\thanks{}
\affiliation{Clarendon Laboratory, Department of Physics, University of Oxford,\\
Parks Road, Oxford, OX1 3PU, \\
United Kingdom.}


\date{\today}

\begin{abstract}
We have measured the precession frequency of a vortex lattice in a
Bose-Einstein condensate of Rb$^{87}$ atoms. The observed mode
corresponds to a collective motion in which all the vortices in
the array are tilted by a small angle with respect to the $z$-axis
(the symmetry axis of the trapping potential) and synchronously
rotate about this axis. This motion corresponds to excitation of a
Kelvin wave along the core of each vortex and we have verified
that it has the handedness expected for such helical waves, i.e.
precession in the opposite sense to the rotational flow around the
vortices.  The experimental method used to excite this collective
mode closely resembles that used to study the scissors mode and
excitation of the scissors mode for a condensate containing a
vortex array was used to determine the angular momentum of the
system. Indeed, the collective tilting of the array that we have
observed has previously been referred to as an `anomalous'
scissors mode.
\end{abstract}

%\keywords{}

\maketitle

Since a vortex was first nucleated in a dilute Bose-condensed gas
\cite{Matthews1999b}, there has been a considerable effort to
understand the dynamical behaviour of individual vortices and
vortex arrays. This research has, in part, been driven by the
parallels between dilute gas systems and more complex superfluids
such as He$^{4}$, however with the Bose-condensed gases it has
proved to be straightforward to obtain images of vortices and to
measure properties of a single vortex. The precession of a single
vortex has been investigated theoretically
\cite{Svidzinsky2000b,Svidzinsky1998a}, and studied experimentally
in a nearly spherical Bose-condensate \cite{Anderson2000a}. It is
also possible to nucleate many vortices in a Bose-condensed gas
and these form a regular Abrikosov lattice
\cite{Madison2000a,Haljan2001a,Raman2001a}. The work described
here can be described in two complementary ways: (a) as an
extension of the previous work on the precession of a single
vortex to the case of an array of vortices in an anisotropic trap
where the collective motion of the vortices is relatively rapid,
or (b) as the excitation of the lowest-energy Kelvin wave of a
vortex lattice. This second viewpoint is described in more detail
below.

\begin{figure}
\includegraphics[width = \columnwidth,height = 5cm]{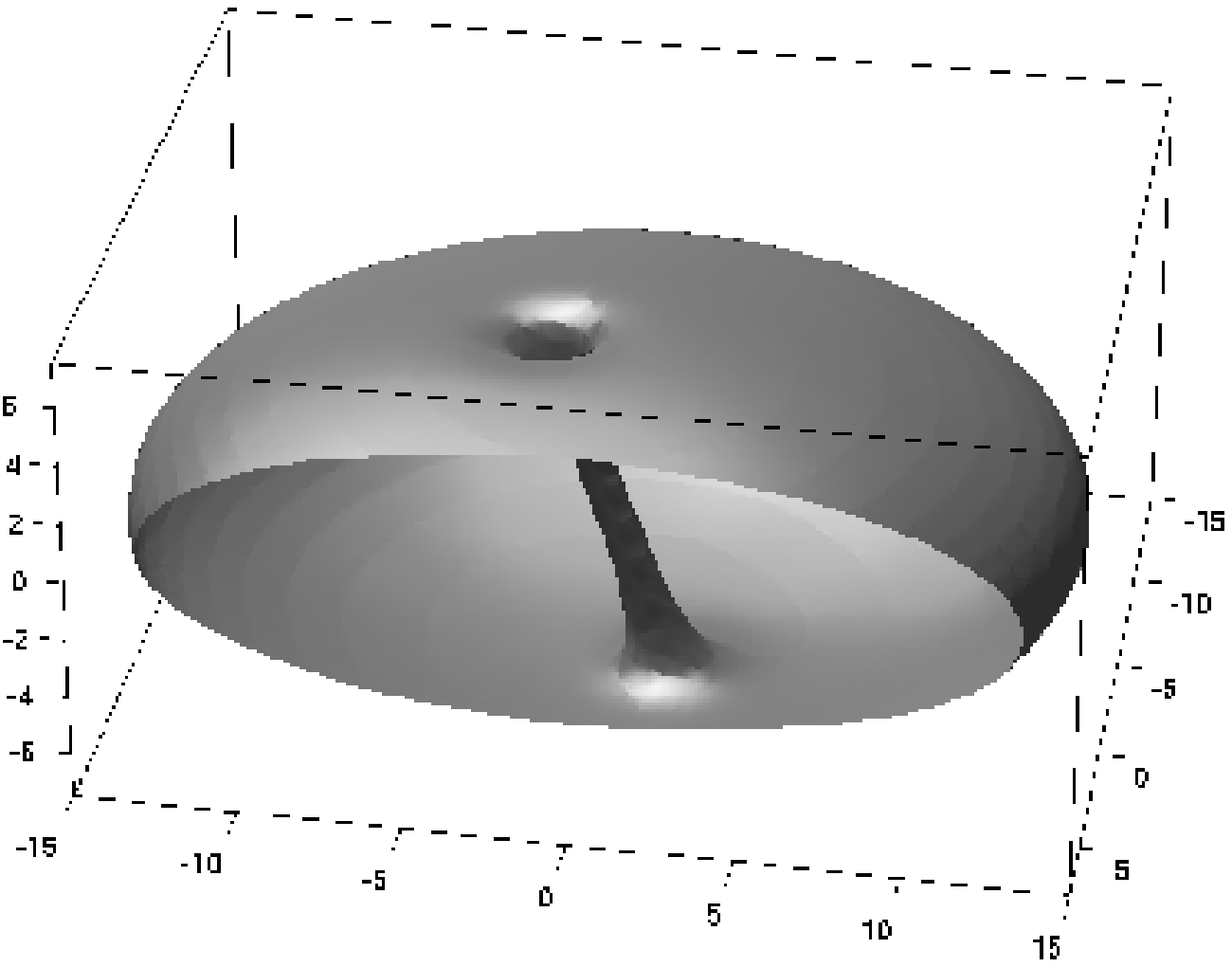}
\includegraphics[width = \columnwidth,height = 5cm]{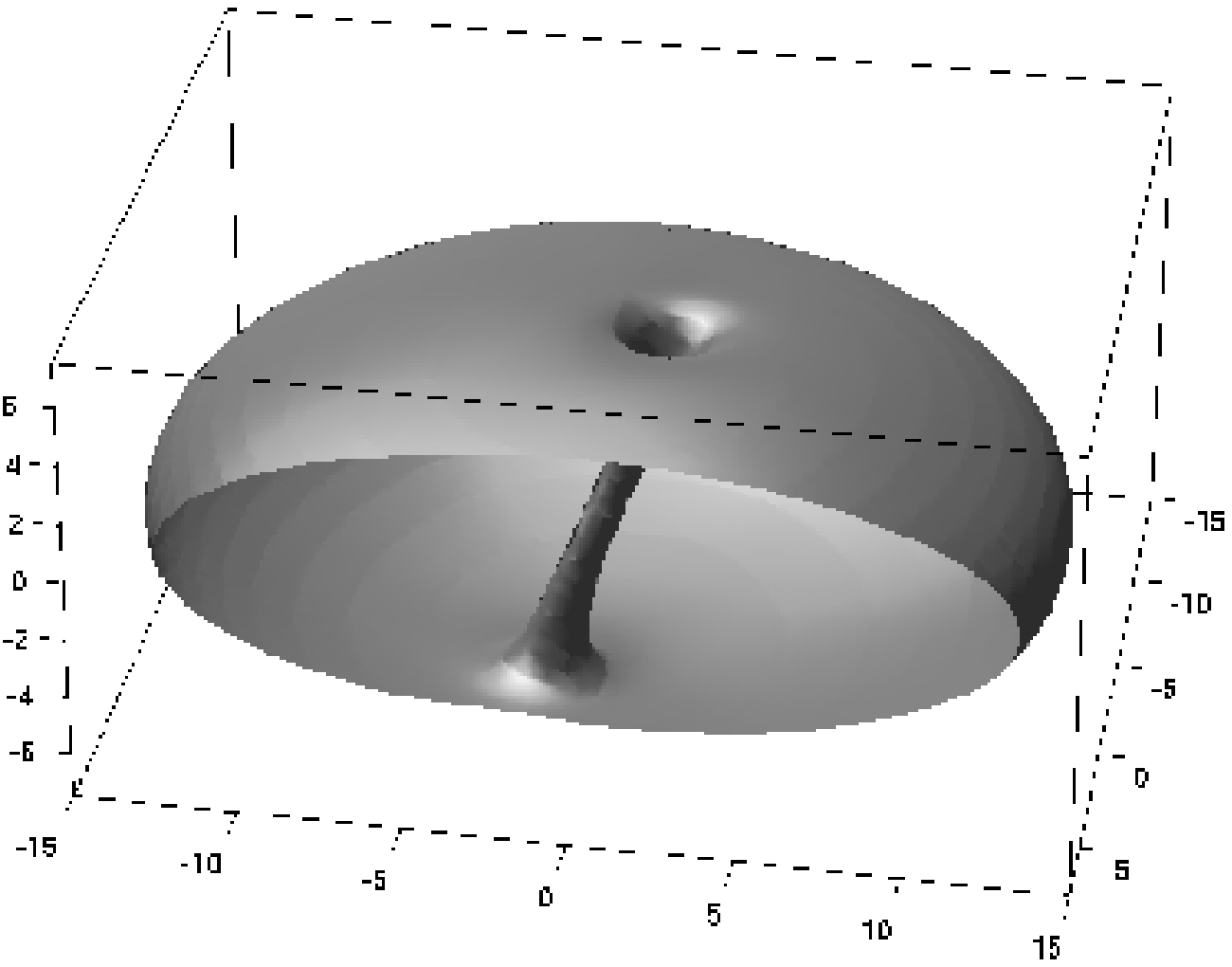}
\includegraphics[width = \columnwidth,height = 5cm]{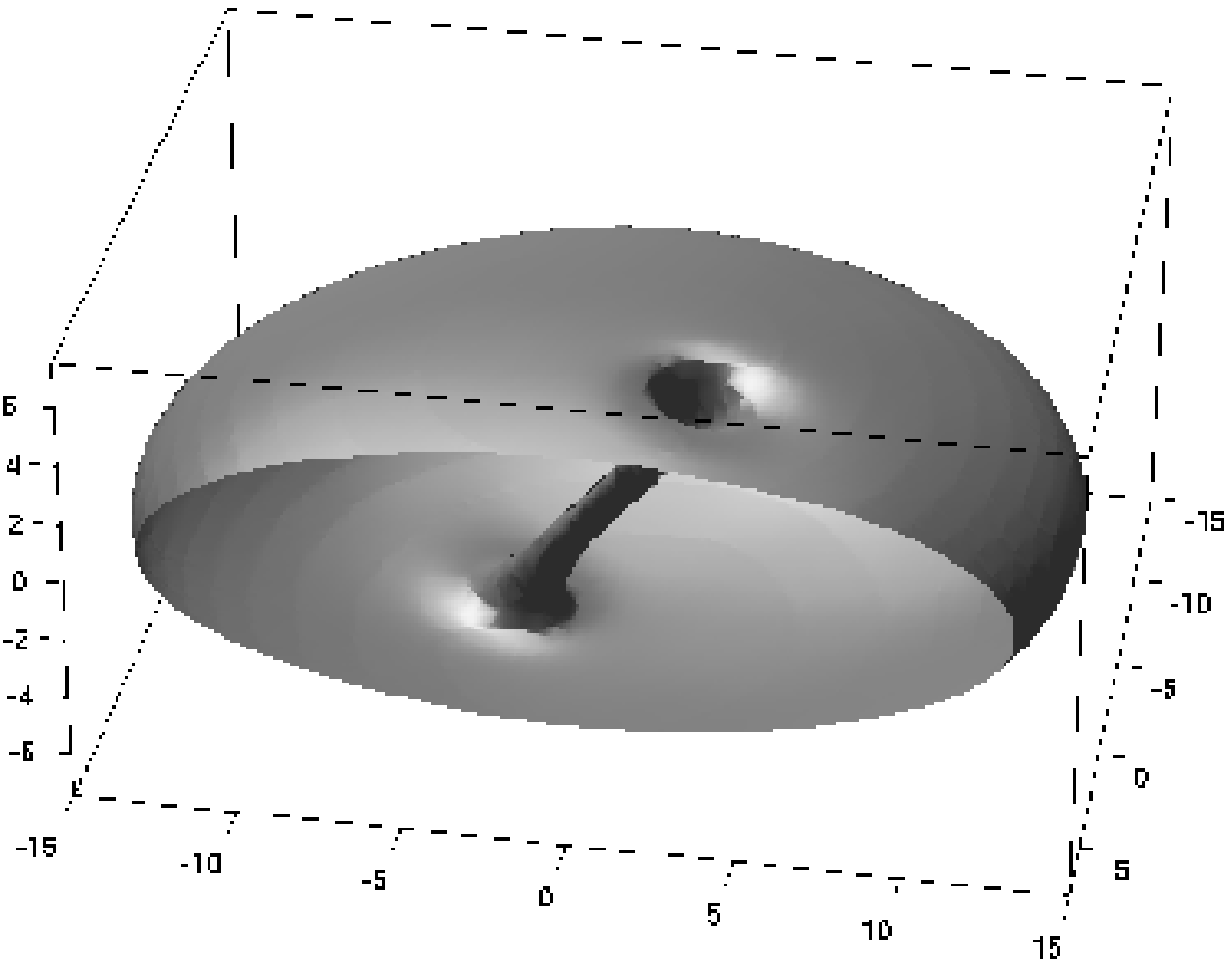}
\includegraphics[width = \columnwidth,height = 5cm]{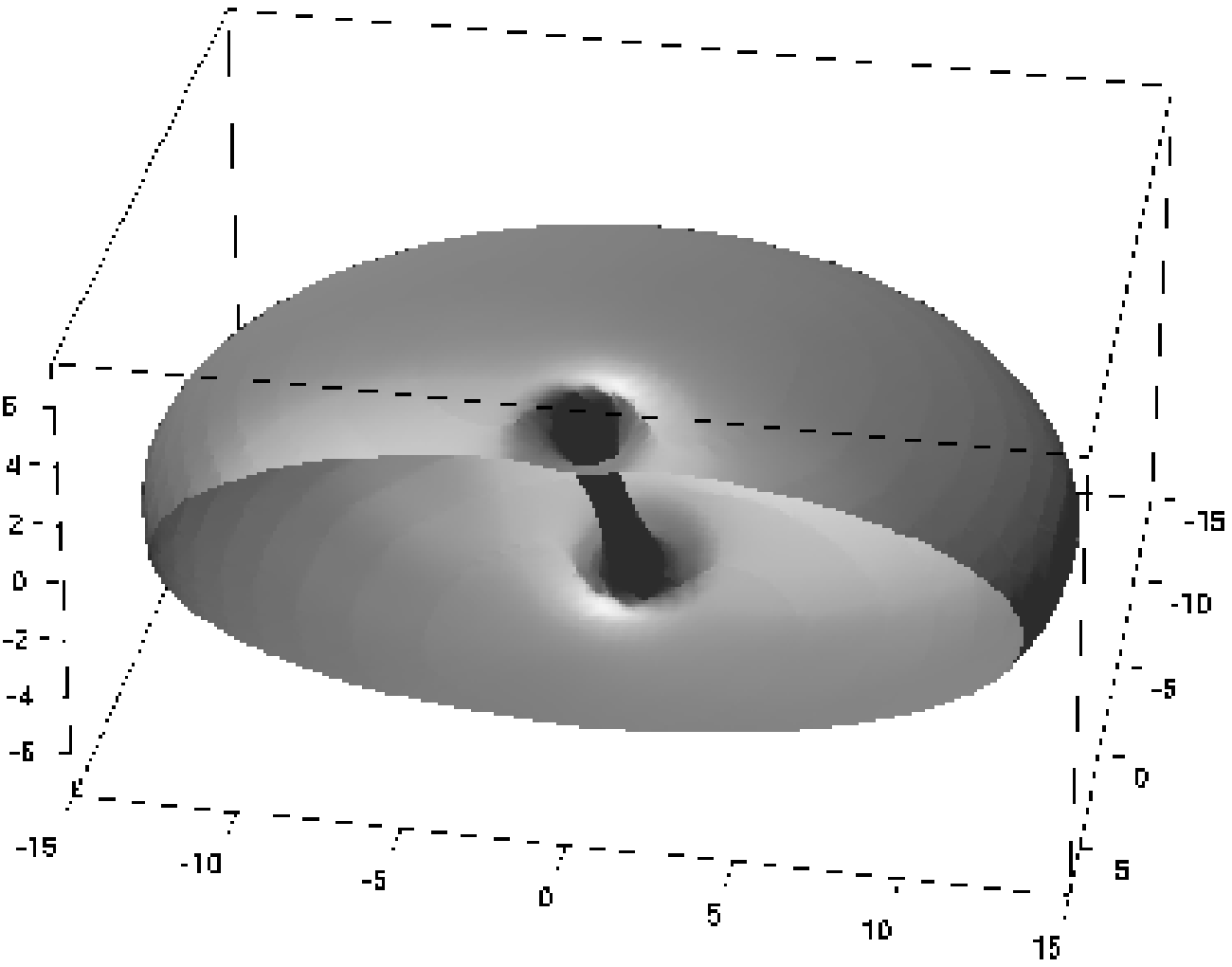}
\caption{The lowest-energy Kelvin wave of a single vortex
corresponds to a rotation of the tilted vortex line in the
opposite direction to the rotation of the condensate. The plots
show one complete rotation of the vortex going from top to bottom.
In the tilting mode of a vortex array excited in this experiment
each vortex synchronously undergoes a similar motion. The plots
show surfaces of constant density $|\Psi|^2$=9e-4 of a condensate
cloud (wave function normalized to one) from a numerical
simulation using the Gross-Pitaevskii equation. The initial state
is a single centred vortex (aligned along the vertical axis); this
is excited by resonant driving for two vortex precession periods
by tilting the confining potential about an axis in the horizontal
plane, as in the experiments, with an amplitude of 4$^{\circ}$. In
this simulation $N = 75000, f_{\bot} = 61$~Hz$, f_{z} = 172.5$~Hz.
}\label{kelvinplot}
\end{figure}

Vortices break the degeneracy of certain modes in the normal
Bogoliubov excitation spectrum for a trapped condensate. This
splitting has been observed for both the m=2 quadrupole mode
\cite{Chevy2000a} and also for the scissors mode
\cite{Hodby2003a}; in the latter case the precession that arises
when the condensate has some angular momentum leads to a
`superfluid gyroscopic' motion \cite{Stringari2001a}. More
recently an excitation of a vortex in the form of a helical Kelvin
wave has been detected \cite{Bretin2003a}; collective oscillations
of a vortex lattice called Tkachenko modes have been observed and
their frequency measured \cite{Coddington2003a}. In a recent
theoretical paper Chevy and Stringari \cite{Chevy2003a} have
extended the hydrodynamic theory of a Bose-condensed gas to
describe an array of vortices and they predict that an array
precesses in a similar way to a single vortex, i.e. all the
vortices simultaneously undergo a motion equivalent to the
lowest-energy Kelvin wave of a single vortex (see
Fig.~\ref{kelvinplot}). (In these waves the orientation of the
vortex core changes and, unlike the Tkachenko modes, the Kelvin
waves do not arise in a two-dimensional system.) In this paper we
present measurements of the frequency of this `collective tilting'
mode of the vortex array by an experimental method similar to that
used to study the scissors mode of the condensate
\cite{Marago2000a}.  Indeed Chevy and Stringari \cite{Chevy2003a}
refer to the collective tilting of the array as an `anomalous'
scissors mode. Their hydrodynamic theory predicts that the
frequencies of this mode and the two normal scissors modes, in a
reference frame rotating at frequency $\Omega_{0}$, are given by
the roots of the cubic equation:
\begin{equation}\label{1}
  \omega^{3}\pm2\Omega_{0}\omega^{2}-\omega(\omega_{z}^{2}+\omega_{\bot}^{2}-\Omega_{0}^{2})\mp2\Omega_{0}\omega_{z}^{2}
= 0,
\end{equation}
where $\omega_{\bot}$ and $\omega_{z}$ are the radial and axial
angular oscillation frequencies. $\Omega_{0} = n_{v}h/2m$ is the
effective rotation frequency of the condensate; $n_{v}$ is the
number of vortices per unit area and m is the mass of the
condensed isotope. The hydrodynamic theory assumes a uniform
distribution of vortices within the condensate so that it mimics
the rotation of a rigid body.

\begin{figure}
\includegraphics[width=\columnwidth]{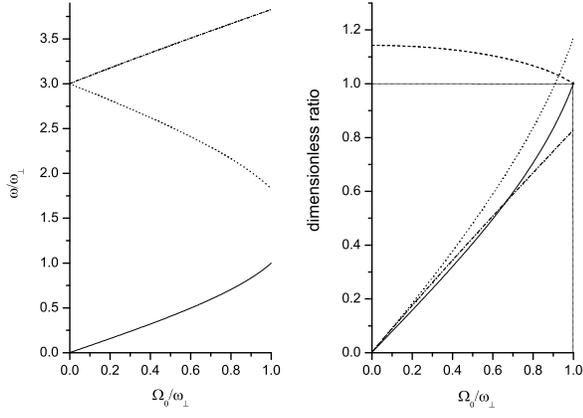}\\
\caption{Calculation of the mode frequencies predicted by Eq.
\ref{1} for a trapping potential with
$\omega_{z}/\omega_{\bot}=\sqrt{8}$. (a)
$\omega_{u}/\omega_{\bot}$ (dot-dash line),
$\omega_{l}/\omega_{\bot}$ (dotted line),
$\omega_{t}/\omega_{\bot}$ (solid line). (b)
$\omega_{u}/\omega_{\bot}-3$ (dot-dash line),
$3-\omega_{l}/\omega_{\bot}$ (dotted line),
$\omega_{t}/\omega_{\bot}$ (solid line). The plot also shows the
ratio $(\omega_{u} - \omega_{l})/2\omega_{t}$ (dashed
line).}\label{freqplotcomb}
\end{figure}

We denote the (angular) frequencies of the upper and lower
scissors modes as $\omega_{u}$ and $\omega_{l}$ respectively and
the frequency of the tilting mode by $\omega_{t}$. These
frequencies are calculated in the non-rotating, laboratory
reference frame. Generally we shall express frequencies in terms
of their fraction of the radial frequency of the trap. The three
solutions of Eq.~\ref{1} in a potential with
$\omega_{z}/\omega_{\bot}=\sqrt{8}$ are shown in
Fig~\ref{freqplotcomb}~(a). This shows that when $\Omega_{0}=0$
the scissors mode frequency is three times the radial trap
frequency, and that the splitting between the upper and lower
scissors modes is about $2\Omega_{0}\simeq 2\omega_{t}$.
Fig.~\ref{freqplotcomb}~(b) shows a plot of the three quantities
$\omega_{u}/\omega_{\bot}-3$, $3-\omega_{l}/\omega_{\bot}$,
$\omega_{t}/\omega_{\bot}$ and also the ratio $(\omega_{u} -
\omega_{l})/2\omega_{t}$. This shows that the frequencies of all
the modes vary approximately linearly with $\Omega_{0}$ (angular
momentum); in particular $\omega_{u}$ is very close to linear over
the entire range but both the lower scissors mode and the tilting
mode show a noticeable deviation. The given ratio has a value
within 15$\%$ of unity throughout the range. Note that the
frequency of the tilting mode tends to zero as $\Omega_{0}$ tends
to zero indicating that this is a mode of the vortices themselves,
in contrast the normal scissors modes are modes of the condensate
whose frequency is perturbed by the presence of vortices (angular
momentum). It can be seen directly from the cubic equation that
for a condensate in a spherically symmetric potential
$(\omega_{x}=\omega_{y}=\omega_{z})$ the `tilting' mode has zero
frequency; this arises because vortices have the same energy for
any orientation in a spherical cloud and so if the orientation of
a vortex is changed with respect to some axis then it will simply
remain at the new angle.

\begin{figure}
\includegraphics[width = 6cm]{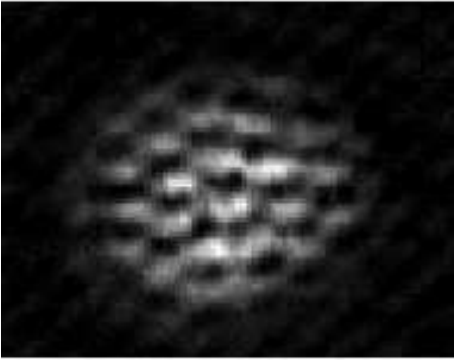}\\
\caption{Absorption image of a vortex array, taken along $z$
axis.}\label{vortices}
\end{figure}

Our experiment uses evaporative cooling in a TOP (Time-averaged
Orbiting Potential) trap to create a Bose-condensate that contains
$\sim10^{5}$ atoms of the Rb$^{87}$  isotope in the $|F=1,
m_{F}=-1 \rangle$ state. The oscillation frequencies of atoms in
the trap are 61~Hz radially and 172.5~Hz along the $z$ (vertical)
direction. After a condensate has been formed we rotate the
trapping potential around the $z$-axis at a frequency of 46.5~Hz,
and adiabatically change from a cylindrical trapping potential to
a rotating elliptical potential over a period of 200~ms. The
potential is rotated  for 700~ms at a final ellipticity where
$w_{x}/w_{y}=0.95$, before it is ramped  back to an
axially-symmetric (circular) potential. During this process
vortices nucleate at the edge of the condensate and then move
towards the centre of the trap \cite{Hodby2001a}. This procedure
introduces ten or more vortices into the condensate
(Fig.~\ref{vortices}).

\begin{figure}
\includegraphics[width = \columnwidth]{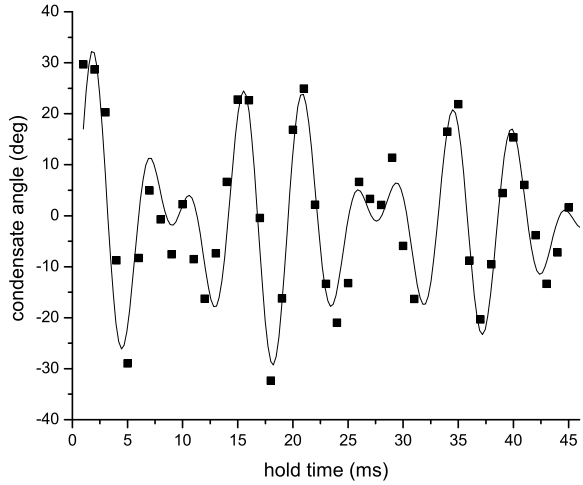}\\
\caption{Exciting the normal scissors mode in the presence of
vortices produces a precession that can be used to infer the
angular momentum of the vortex array.}\label{gyrodata}
\end{figure}

The trapping potential is tilted by the addition of an oscillating
field in the $z$-direction, in phase with the usual rotating field
in the TOP trap. When no vortices are present, tilting the
trapping potential excites the scissors mode as described in
\cite{Marago2000a}. This scissors motion is the superposition of
two degenerate modes characterized by functions of the form
$f(r)xz$ and $f(r)yz$ \cite{Ohberg1997a}. The presence of one or
more vortices, breaks the degeneracy and leads to modes described
by $f(r)z(x \pm \rm{i}y)$; these are eigenstates of the
z-component of orbital angular momentum $L_{z}$. A sudden tilt of
the trapping potential excites a superposition of these two
counter-rotating modes and the resulting scissors oscillation
precesses at a rate equal to the frequency splitting between the
two modes divided by two \cite{Hodby2003a,Stringari2001a}. This
motion is illustrated in Fig.~\ref{gyrodata} for a condensate that
contains an array of many vortices. In this case the effective
rotation frequency of the condensate $\Omega_{0}$ is much greater
than that for a single vortex. A fit to these data gives the
frequencies of the upper and lower scissors modes as $f_{u}$ =
$\omega_{u}/2\pi$ = 211.8 $\pm$ 2.0~Hz and $f_{l}$ = 156.3 $\pm$
2.0~Hz respectively. We find that this corresponds to an average
angular momentum per particle of $\langle l_{z} \rangle=8.4 \pm
0.4~\hbar$, for a total number of atoms $N = 75\,000$. A vortex at
the trap centre contributes $\hbar$ of angular momentum per
particle but off-centre vortices have a smaller contribution
\cite{Svidzinsky2000b}. From these measured frequencies and
Eq.~\ref{1}, an effective rotation of $\Omega_{0} = 0.495 \pm
0.019$ can be deduced, which implies a tilting mode frequency
$\omega_{t}$ = 23.5 $\pm$ 0.9~Hz according to the hydrodynamic
theory.

\begin{figure}
\includegraphics[width = \columnwidth]{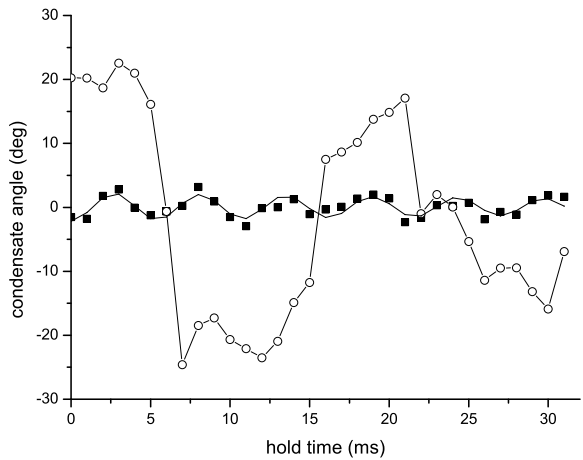}\\
\caption{(black squares) Off resonant excitation of the scissors
mode when no vortices are present. (white circles) Resonant
driving of the `tilting mode' in the presence of a vortex
array.}\label{anomcontrol}
\end{figure}

\begin{figure}
\includegraphics[width = \columnwidth]{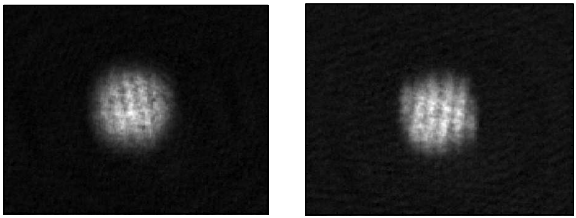}\\
\caption{Absorption images of the vortex lattice, viewed in a
radial direction.}\label{tiltingvtx}
\end{figure}

\begin{figure}
\includegraphics[width = \columnwidth]{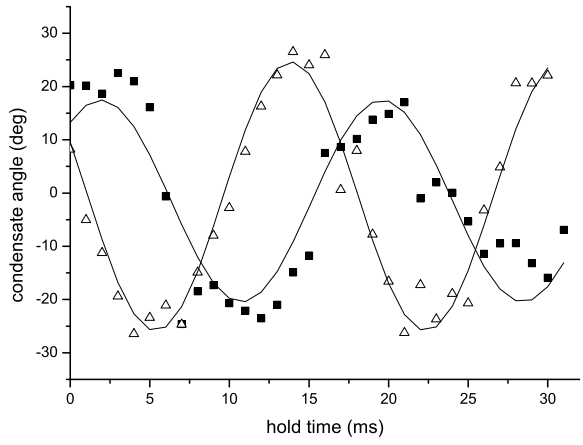}\\
\caption{(black squares) Anomalous mode initially excited
perpendicular to probe beam. (white triangles) Anomalous mode
initially excited parallel to probe beam.}\label{anom2dir}
\end{figure}

To excite the tilting mode of the vortex array the tilt angle of
the trapping potential (relative to the $z$ axis) was driven at a
frequency of 61~Hz for two complete cycles of oscillation with an
amplitude of 0.07~radians. After excitation, the condensate was
held in the trap for a variable amount of time during which the
tilting mode evolved at its natural frequency; the condensate was
then released and allowed to expand freely for 19~ms before a
laser beam was flashed on to record an image of the cloud. (This
imaging was destructive.) When no vortices were present in the
condensate the response to the driving was an excitation of the
normal scissors mode with a small amplitude, as in
Fig.~\ref{anomcontrol}, because this mode is far from resonance.
For a condensate containing vortices, however, there was
near-resonant driving of the tilting mode, which resulted in large
amplitude oscillations of this mode. The tilting of the vortex
array leads to a change in angle of the condensate and the
projection of this motion onto the imaging direction gives a
signal similar to that of the normal scissors mode but at a much
lower frequency. In a random sample of the absorption images the
vortex cores line up with the imaging beam: in these images it is
possible to see that the vortices are indeed tilting in unison
when the mode is excited (see Fig.~\ref{tiltingvtx}). The tilting
mode was excited both along the direction of the imaging beam and
in a direction perpendicular to it (Fig.~\ref{anom2dir}). This
allowed us to determine the direction of rotation of the tilting
mode, and to verify that it rotates in the opposite direction to
the initial rotation that creates the vortices (the initial
rotation has the same direction as the flow around the vortices).
In a Kelvin wave the vortex core has the form of a helix of a
particular handedness with respect to the direction of rotation of
the vortex; this property also applies to the collective tilting
of the vortex array. A sinusoidal fit to the data gives a
frequency of 57.7 $\pm$ 1.3~Hz for the tilting mode.

The number and position of the vortices varies from shot to shot
resulting in slightly different frequencies of the tilting mode so
that after a few cycles the observations have more fluctuations
(i.e.\ this method of recording the data has a dephasing analogous
to the transverse relaxation in magnetic resonance techniques, and
we have not measured the damping of this mode equivalent to
longitudinal relaxation time.) The uncertainty produced by
variations in the initial conditions could be reduced by taking
more data for each evolution time and averaging, as in previous
work on the superfluid gyroscope \cite{Hodby2003a}.

Using the measured values of $f_{u}$, $f_{l}$ and $f_{t}$ we find
the ratio of frequencies $(f_{u}-f_{l})/(2f_{t}) = 27.7/57.8 =
0.48 \pm 0.03$ that is not consistent with predictions of the
hydrodynamic theory, shown in Fig.~\ref{freqplotcomb}~(b). If we
assume that the scissors modes have the correct frequencies, then
the measured value for the tilting frequency is about 2.5 times
the hydrodynamic prediction. We should emphasize, however, that we
cannot determine from the frequency measurements whether the
scissors modes are more accurately described by the theory than
the tilting mode. One way of checking this is to estimate the
contributions to the angular momentum from the number of observed
vortices and their position within the cloud, and compare this
with the value obtained from the scissors mode splitting: a vortex
at radius $r$ in a condensate of Thomas-Fermi radius $R$
contributes $N\hbar(1-r^{2}/R^{2})^{5/2}$ to the total angular
momentum. In this way we found a total angular momentum per
particle of 8.5 $\pm$ 1.0~$\hbar$ for the image in
Fig.~\ref{vortices}, which is in good agreement with the value
calculated from the scissors mode frequency.

If the prediction of the scissors mode frequency is correct, then
we need to explain the difference between the predicted and
measured values of the tilting mode frequency. Firstly, to
eliminate the possibility that we were exciting a higher order
Kelvin mode, the tilt angle of the trapping potential was driven
at 25~Hz after a vortex array had been nucleated. This driving
frequency is close to the hydrodynamic prediction, but no response
from the condensate was observed. If we are definitely driving the
lowest order mode, the case for which is supported by the
straightness of the vortex cores in Fig.~\ref{tiltingvtx}, then
the difference is most likely explained by considering the range
of applicability of the hydrodynamic theory; indeed, the theory
relies on the fact that it is possible to average physical
quantities over domains containing several vortices
\cite{Chevy2003a}. For a single vortex a full numerical simulation
is currently possible, as shown in Fig.~\ref{kelvinplot} for our
experimental conditions. The tilting mode frequency is calculated
to be $f_{t}= 0.27\omega_{z}$, where $\omega_{z}$ is the axial
trapping frequency; the splitting between the scissors modes is
calculated to be $0.004\omega_{z}$, thus for a single vortex
$(f_{u}-f_{l})/(2f_{t}) = 0.09$. This deviates even further from
the hydrodynamic prediction that the frequency ratio is about
unity than the results for our small vortex array. The
hydrodynamic theory only gives an accurate description for arrays
that contain a large number of vortices, the vortex arrays in this
experiment do not contain a sufficient number of vortices to be in
the regime where it is possible to average over domains of
vortices and apply the hydrodynamic theory (Fig.~\ref{vortices});
increasing the value of experiment in this intermediate region.
The hydrodynamic theory should, however, be a better approximation
for larger vortex arrays as in references
\cite{Haljan2001a,Raman2001a}.

In conclusion, this experimental measurement of the frequency and
direction of rotation of the collective tilting mode of the vortex
array qualitatively supports the predictions made by Chevy and
Stringari \cite{Chevy2003a}. The cause of the discrepancy between
our measurements and the hydrodynamic predictions could be
investigated experimentally with a larger condensate that can
contain a higher number of vortices or theoretically by a
numerical simulation as in \cite{Nilsen2002a}. The experimental
method for direct excitation of a vortex (or vortex array) can
also be used to study higher-order Kelvin waves.

\begin{acknowledgments}
The authors would like to acknowledge financial support from the
EPSRC and ARDA.
\end{acknowledgments}


\end{document}